\documentclass[nofootinbib,superscriptaddress,longbibliography,a4paper,twocolumn]{revtex4-1}
\usepackage{graphicx}
\usepackage{dcolumn}
\usepackage{bm,bbm}
\usepackage{xcolor}
\usepackage{tcolorbox}
\usepackage{algorithm}
\usepackage{algpseudocode}
\usepackage{amsmath}
\usepackage{bbold}
\usepackage{braket}
\usepackage{nameref}
\usepackage[toc,page]{appendix}

\usepackage[colorlinks,breaklinks,linkcolor={blue},citecolor={magenta},urlcolor={blue}]{hyperref}

\makeatletter
\newcommand\org@hypertarget{}
\let\org@hypertarget\hypertarget
\renewcommand\hypertarget[2]{%
  \Hy@raisedlink{\org@hypertarget{#1}{}}#2%
  }
\makeatother

\DeclareMathOperator{\Tr}{Tr}

\usepackage{amsthm}

\newtheorem{Definition}{{\bf Definition}}
\newtheorem{Lemma}{{\bf Lemma}}

\begin{document} 

\title{Accessing inaccessible information via quantum indistinguishability}

\author{Sebastian Horvat}
\email{sebastian.horvat@univie.ac.at}
\affiliation{University of Vienna, Faculty of Physics {\&} Vienna Doctoral School in Physics {\&} Vienna
Center for Quantum Science and Technology, Boltzmanngasse 5, 1090 Vienna, Austria}

\author{Borivoje Daki\'c}
\email{borivoje.dakic@univie.ac.at}
\affiliation{University of Vienna, Faculty of Physics, Vienna Center for Quantum Science and Technology, Boltzmanngasse 5, 1090 Vienna, Austria}
\affiliation{Institute for Quantum Optics and Quantum Information (IQOQI), Austrian Academy of Sciences, Boltzmanngasse 3, 1090 Vienna, Austria.}

\begin{abstract}
In this paper we present and analyze an information-theoretic task that consists in learning a bit of information by spatially moving the ``target'' particle that encodes it. We show that, on one hand, the task can be solved with the use of additional independently prepared quantum particles, only if these are indistinguishable from the target particle. On the other hand, the task can be solved with the use of distinguishable quantum particles, only if they are entangled with the target particle. Our task thus provides a new example in which the entanglement apparently inherent to independently prepared indistinguishable quantum particles is put into use for information processing. Importantly, a novelty of our protocol lies in that it does not require any spatial overlap between the involved particles. Besides analyzing the class of quantum-mechanical protocols that solve our task, we gesture towards possible ways of generalizing our results and of applying them in cryptography.
\end{abstract}

\date{\today}
\maketitle
\section{Introduction}

Indistinguishable particles are commonly represented in quantum-mechanical models in two distinct ways: roughly, the first-quantization formalism models their state vectors as elements of the (anti)symmetric subspace of a multi-particle Hilbert space, whereas the second-quantization formalism takes their state space to be given by a Fock space of modes.\footnote{Note that the terms ``indistinguishable particles'' and ``identical particles'' are sometimes used to refer to different things in the literature (see e.g. \cite{her2001, her2006}). Nevertheless, these terms will be used synonymously in the present article.} Interestingly, the former option allows for the state of independently prepared indistinguishable particles - e.g. the state of two electrons stemming from causally disconnected galaxies - to be entangled, despite the absence of common past interactions between the particles. On the contrary, in the second-quantization formalism, the entanglement between the particles' modes is contingent on the particles' past interactions: indeed, the state of independently prepared particles is represented as separable.

This representational ambiguity motivates the question of whether independently prepared indistinguishable particles ought to be regarded as entangled or not: possible answers are still being actively debated \cite{ghirardi2002entanglement, ghir2003, ghirardi2004general, johann2021locality, benatti2021entanglement, benatti2020entanglement}. Furthermore, since entanglement is a paradigmatic resource for quantum information processing, another question that has arisen is whether the entanglement that is apparently intrinsic according to the first-quantization formalism can in principle be accessed and used: here an affirmative answer has been supported by various tasks in which particle indistinguishability provides similar advantages to those obtainable from entanglement \cite{li2001, pas2001, schl2001, eck2002, gitt2002, zan2002, omar2002, paun2002, shi2003, ved2003, wise2003, bose2003, sher2006, omar2006, cavalcanti2007useful, tichy2013, krenn2017, li2017, franco2018indistinguishability, chin2019, kar2019, ju2019, barghathi2019operationally, castellini2019, castellini2019b, nosrati2020a, nosrati2020b, morris2020entanglement, barros2020, sun2020, holmes2020, chin2021, wang2022, zaw2022, sun2022, lee2022}. A feature that is common to these tasks is that the advantage brought forth by particle indistinguishability is present only if the involved particles spatially overlap at least at some moment during the protocol, i.e. the particles should not be fully distinguishable by their spatial degree of freedom, if their entanglement is to be exploitable.

In this paper we propose a new information-theoretic task that can be solved with the use of independently prepared indistinguishable particles, but cannot be solved with distinguishable ones, unless they are entangled. We thereby show that, in the context of our task, the indistinguishability of independently prepared particles provides an equal resource to the one obtainable from the entanglement of distinguishable particles, thus providing further support to the claim that the entanglement inherently present according to the first-quantization formalism is more than a mere mathematical artefact, and can indeed be used in information processing. Unlike previously proposed examples that support the latter claim, our protocol does not require its involved particles to spatially overlap at any moment - in fact, the protocol's success does not depend in any way on the distance between the particles.

The paper is structured as follows. Section \ref{sec1} offers an informal presentation of our task, whereas its precise formulation is given in Section \ref{sec2}. The same section also contains the quantum-mechanical analysis of the task and an expansion on the requirements necessarily satisfied by any quantum-mechanical protocol that solves it. Finally, Section \ref{sec3} provides a discussion and an outlook on possible future developments of our results and on their potential practical applications.

\section{Informal presentation}\label{sec1}
\begin{figure}
\includegraphics[width=\linewidth]{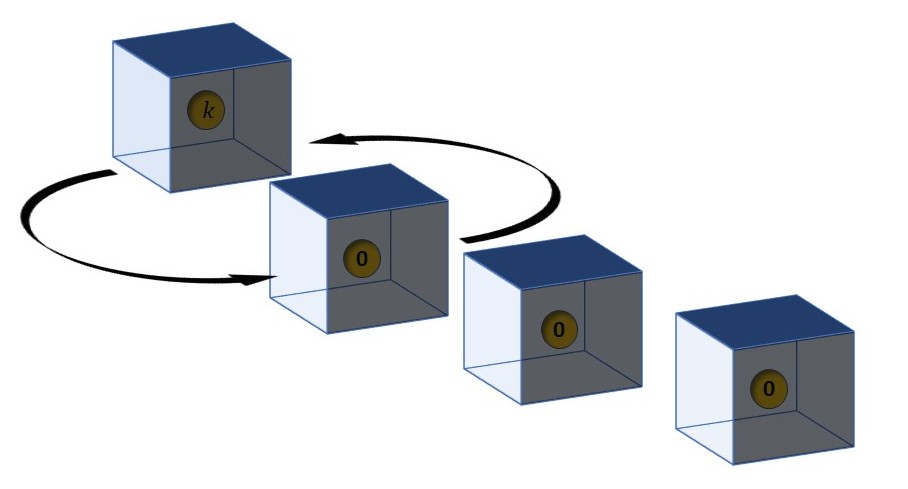}
\caption{Alice is given a box containing an object prepared in state $k\in \left\{0,1\right\}$. She also has at disposal other boxes, with their pertaining objects set in state $0$. Alice is challenged to learn the value $k$ by solely moving the box that contains the hidden object (without opening it), while also being allowed to implement transformations on the other boxes: for example, as represented in the diagram, she can swap the locations of the first two boxes.}
\centering
\label{fig1}
\end{figure}

Consider the following scenario. An agent named Alice is given a box, which contains a physical object (e.g. an atom) prepared in one of two perfectly distinguishable states, thereby encoding one bit of information. The box is sealed in such a way that Alice does not have the means of opening it and thus cannot access the object directly. Suppose that the only action she can perform on the box is to move it in space, e.g. she can move the box from some initial position $\vec{x}$ to another position $\vec{x}'$. Alice may also have available various experimental devices and other physical objects (as for example other boxes). Lastly, we will assume that spatially moving the box leaves the state of the object within it invariant, and that Alice cannot infer the state by merely moving the box (e.g. via some state-dependent back-reaction acting on Alice's experimental devices or additional objects). An example of this scenario, where the additional objects consist of other boxes, is pictured in Fig. \ref{fig1}.\\ 
Now we are ready to formulate the task: \textit{can Alice learn the state of the object within the box, given the above restrictions on what she is allowed to do?} At the current stage, this question may appear nonsensical: if (i) Alice is only able to move the box, and (ii) a mere movement of the box cannot reveal the state of the object within it, then it trivially follows that Alice cannot accomplish the task! However, if the box can be treated quantum-mechanically and if Alice has at disposal additional identical quantum systems, then, as we will show in the next section, she can in fact learn the state of the object, while - in a sense that will be made precise below - still only moving the box. Let us now provide a formalized version of the task and the quantum-mechanical protocol that can be used to solve it.

\section{Formalization of the task}\label{sec2}
Here we will put the scenario presented in the previous section on more formal and abstract grounds. Consider that Alice is given a localized physical object, henceforth named $\cal{T}$. $\cal{T}$ can be characterized to have two degrees of freedom (d.o.f.): an intrinsic one, which we will label with $k$, and a position, labelled with $\vec{x}$. We will thus represent the overall state of the object with the ordered pair $(\vec{x},k)_T$. Throughout the paper, we attach only an \textit{operational meaning} to states, i.e. the sentence ``the object is in state $(\vec{x},k)_T$'' is hereafter synonymous with ``a position measurement on the object \textit{would} output value $\vec{x}$, and an appropriate measurement of the internal d.o.f. \textit{would} output value $k$''. We furthermore assume that $k$ can take only two possible values, 0 or 1. Let us now suppose that Alice is only allowed to move the object through space, while keeping the internal d.o.f. intact. We will model this restriction by assuming that she has at disposal a device $\cal{D}$ that can implement any transformation in $\mathbf{M}$, where $\mathbf{M}$ is a set of transformations $M: \mathbb{R}^3 \rightarrow \mathbb{R}^3$, the latter being operators that map position vectors $\vec{x}$ into position vectors $\vec{x}'=M\vec{x}$. We will associate states $(M)_D$ to the device, where $M$ indicates which map the device is set to implement on a potential target object. The device is constructed in such a way that the dynamical interaction with the localized target object $\cal{T}$, which is initially in state $(\vec{x},k)_T$, is given by
\begin{equation}\label{eq1}
    (M)_D(\vec{x},k)_T \rightarrow (M)_D(M\vec{x},k)_T.
\end{equation}

Furthermore, Alice may have at disposal other localized objects, which cannot interact directly with the initial object $\cal{T}$, but can interact with the device $\cal{D}$. Labelling collectively these additional objects with $\cal{A}$, and their pertaining state with $(\alpha)_A$, the latter constraint means that the overall interaction between $\cal{D}$, $\cal{A}$ and $\cal{T}$ can be represented as:
\begin{equation}\label{eq11}
    (M)_D(\alpha)_A(\vec{x},k)_T \rightarrow (M)_D(\tilde{\alpha}(M,\alpha))_A(M\vec{x},k)_T,
\end{equation}
where the final state $\tilde{\alpha}(M,\alpha)$ of $\cal{A}$ is in principle any function insensitive to the internal d.o.f. pertaining to $\cal{T}$. Moreover, if the initial state of $\cal{D}$ is not equal to any $(M)_D$ (but is e.g. a probabilistic mixture, a quantum superposition state or some generalized probabilistic state \cite{barrett2007information, plavala2021general}), we assume that it is still the case that, were one to find the final state of $\cal{D}$ to be $(M)_D$, one would find the final state of $\cal{T}$ and $\cal{A}$ to be $(\tilde{\alpha}(M,\alpha))_A(M\vec{x},k)_T$. Stated more precisely, for any initial state of $\cal{D}$, post-selecting the final state on a definite state $(M)_D$ of $\cal{D}$ results in the postselected joint state being equal to the RHS of Eq. \eqref{eq11}. Finally, we assume that, for arbitrary initial states of $\cal{D}$, the probability of finding $\cal{D}$ in state $(M)_D$ does not change upon the above interaction.

Now suppose that Alice is allowed only to interact with her device, i.e. she is solely able to control and to read out the state of $\cal{D}$. In other words, she can interact only indirectly with objects $\cal{T}$ and $\cal{A}$, via mediation through $\cal{D}$. Furthermore, Alice has perfect knowledge of the initial state $(\alpha)_A$ associated to the additional objects $\cal{A}$, but does not have any prior knowledge of the value $k$ pertaining to $\cal{T}$. The information-theoretic task can now be formulated as follows:\\
\\
\textbf{Task}: \textit{can Alice learn the state $k$ pertaining to $\cal{T}$, solely by manipulating the given device $\cal{D}$?}\\

As can be immediately seen from Eq. \eqref{eq11}, the state of $\cal{D}$ is insensitive to $k$, so the answer to the latter question seems to be an immediate `\textit{No}'. Also, at the current stage, it is admittedly not clear at all what role, if any, the additional objects $\cal{A}$ could play for the accomplishment of the required task. However, in what follows, we will show that if $\cal{D}$, $\cal{A}$ and $\cal{T}$ can be treated as quantum systems, and if Alice is able to coherently manipulate the device $\cal{D}$, then she can in fact learn the required value $k$, given that $\cal{A}$ and $\cal{T}$ can be modelled as indistinguishable quantum particles.

\subsection*{The quantum protocol}
We will now assume that $\cal{D}$, $\cal{A}$ and $\cal{T}$ can be treated as quantum systems that can be modelled via the usual rules of quantum mechanics. Therefore, we can associate a quantum state 
\begin{equation}
    \ket{\psi} \in \cal{H}\equiv H_D\otimes H_A \otimes H_T
\end{equation}
to the joint system comprised of $\cal{D}$, $\cal{A}$ and $\cal{T}$. The Hilbert space $\cal{H}_D$ associated to the device is spanned by vectors $\left\{\ket{M}, \forall M \in \mathbf{M} \right\}$, whereas the Hilbert space $\cal{H}_T$ associated to $\cal{T}$ is spanned by $\left\{\ket{\vec{x},k} \equiv \ket{\vec{x}}\otimes \ket{k}, \forall \vec{x} \in \mathbb{R}^3, \forall k=0,1 \right\}$. Vectors $\ket{M}$ are eigenstates of the observable $\hat{M}$ that corresponds to the ``measurement'' of the device in the intended classical basis, whereas $\ket{\vec{x}}$ and $\ket{k}$ are respectively eigenstates of observables $\hat{\vec{x}}$ and $\hat{k}$ corresponding to measurements of the position and internal d.o.f. of $\cal{T}$. The structure of the Hilbert space $\cal{H}_A$ pertaining to $\cal{A}$ remains for now unspecified. We now assume, in accord with Eq. \eqref{eq11}, that the interaction between $\cal{D}$, $\cal{A}$ and $\cal{T}$ is given by the following unitary evolution:
\begin{equation}\label{eq2}
    \ket{M}_D \ket{\alpha}_A \ket{\vec{x},k}_T \rightarrow \ket{M}_D \ket{\tilde{\alpha}(M,\alpha)}_A\ket{M\vec{x},k}_T.
\end{equation}
Notice that the above corresponds to a control gate, where the device $\cal{D}$ acts as the control system, and the objects $\cal{T}$ and $\cal{A}$ act as targets. We stress that we are assuming that $\cal{D}$, $\cal{A}$ and $\cal{T}$ are not interacting with any further environment, i.e. that they constitute an isolated system; in the next subsection we will analyze protocols that violate this assumption.

Let us now specify the additional objects $\cal{A}$ to consist only of \textit{one} object localized at position $\vec{x}_A$, and possessing a binary internal d.o.f., whose value is for simplicity set to $k_A=0$. The Hilbert space associated to $\cal{A}$ is thus isomorphic to the one associated to $\cal{T}$, and the object $\cal{A}$ is assigned state $\ket{\vec{x}_A,0}$. We also assume that the device $\cal{D}$ can be used to move $\cal{A}$ through space. Consequently, Alice is now able to \textit{swap} the two objects with the use of her device, by moving $\cal{T}$ to position $\vec{x}_A$, and $\cal{A}$ to position $\vec{x}_T$. In order to simplify the discussion, let us introduce an effective state of $\cal{D}$, which we will label with `$S$', and that is constructed in such a way that it effectively swaps the two objects upon interaction, i.e. 
\begin{equation}\label{eq3}
    \ket{S}_D \ket{\vec{x}_A,0}_A \ket{\vec{x}_T,k_T}_T  \rightarrow \ket{S}_D  \ket{\vec{x}_T,0}_A \ket{\vec{x}_A,k_T}_T,
\end{equation}
where we introduced more indices in order to avoid confusion.

Under the assumption that $\cal{A}$ and $\cal{T}$ are distinguishable systems, Eq. \eqref{eq2} implies that Alice cannot accomplish the task, because the internal degree of freedom pertaining to $\cal{T}$ is isolated from the other subsystems. Let us now suppose that objects $\cal{A}$ and $\cal{T}$ are \textit{indistinguishable}, i.e. that they can be modelled as indistinguishable quantum particles: here we assume for simplicity the bosonic case, even though the protocol would also work in the fermionic case, as it can be easily checked. This warrants us to introduce the ``second quantization'' notation via the following recipe: 
\begin{equation}\label{eq4}
    \ket{\vec{x}_A,0}_A \ket{\vec{x}_T,k_T}_T \Rightarrow  \ket{0}_{\vec{x}_A}\ket{k_T}_{\vec{x}_T},
\end{equation}
where the latter state means that at location $\vec{x}_A$ there is an object with internal state $0$, and at location $\vec{x}_T$ an object with internal state $k_T$, with no further labels that may distinguish the objects \footnote{In order to avoid potential confusion, let us clarify the connection between the above ``second quantization'' notation, and the more familiar formalism defined in terms of ladder operators. According the the latter, our two-boson state-space is spanned by vectors $\left\{a^{\dagger}_{\vec{y},i}a^{\dagger}_{\vec{z},j}\ket{\Omega}, \forall \vec{y},\vec{z}=\vec{x}_A,\vec{x}_T, \forall i,j=0,1 \right\}$, where $\ket{\Omega}$ is the vacuum state, and $a^{\dagger}_{\vec{y},i}$ is a bosonic ladder operator that creates a boson at position $\vec{y}$ with internal d.o.f. having value $i$. Since throughout our protocol the two bosons never occupy the same position, we are warranted to restrict our two-boson state-space to the subspace $\left\{a^{\dagger}_{\vec{x}_A,i}a^{\dagger}_{\vec{x}_T,j}\ket{\Omega}, \forall i,j=0,1 \right\}$, which is isomorphic to the joint space of two qubits, i.e. to the space $\mathcal{H}_{\vec{x}_T} \otimes \mathcal{H}_{\vec{x}_A}$ introduced above. Stated explicitly, the isomorphism is given by $\ket{i}_{\vec{x}_A}\ket{j}_{\vec{x}_T}\cong a^{\dagger}_{\vec{x}_A,i}a^{\dagger}_{\vec{x}_T,j}\ket{\Omega}$.}. Mathematically, the quantum state is now an element of $\mathcal{H}_{\vec{x}_T} \otimes \mathcal{H}_{\vec{x}_A}$, where $\mathcal{H}_{\vec{x}_{T/A}}$ is associated to spatial mode $\vec{x}_{T/A}$ and is spanned by vectors $\left\{\ket{0}_{\vec{x}_{T/A}},\ket{1}_{\vec{x}_{T/A}} \right\}$. The interaction between the device $\cal{D}$ set in state `$S$' and the two indistinguishable objects is thus given by:
\begin{equation}\label{eq5}
    \ket{S}_D \ket{0}_{\vec{x}_A} \ket{k_T}_{\vec{x}_T} \rightarrow \ket{S}_D  \ket{k_T}_{\vec{x}_A} \ket{0}_{\vec{x}_T}.
\end{equation}

Now comes the crucial observation: unlike as in the case of distinguishable objects (see Eq. \eqref{eq3}), the overall quantum state in Eq. \eqref{eq5} is invariant upon interaction if and only if $k_T=k_A$, i.e. $k_T=0$. This enables the construction of the following protocol that can be used by Alice to learn $k_T$ with probability higher than $\frac{1}{2}$, given that she knows the value $k_A=0$ of the additional system $\cal{A}$. The procedure goes as follows.\\
(a) Alice prepares the device in state $\ket{\phi} \equiv \frac{1}{\sqrt{2}}\left( \ket{\mathbb{1}}_D + \ket{S}_D \right)$, where `$\mathbb{1}$' is the state of the device that implements the identity transformation.\\
(b) She then lets $\cal{D}$, $\cal{A}$ and $\cal{T}$ interact as:
\begin{equation}\label{eq6}
\begin{split}
    &\frac{1}{\sqrt{2}}\left( \ket{\mathbb{1}}_D + \ket{S}_D \right)  \ket{0}_{\vec{x}_A} \ket{k_T}_{\vec{x}_T} \rightarrow \\
    &\frac{1}{\sqrt{2}}\left( \ket{\mathbb{1}}_D \ket{0}_{\vec{x}_A} \ket{k_T}_{\vec{x}_T}  + \ket{S}_D \ket{k_T}_{\vec{x}_A}\ket{0}_{\vec{x}_T}  \right).
    \end{split}
\end{equation}
After the interaction, the reduced density state $\rho$ associated to $\cal{D}$ is:
\begin{equation}\label{eq7}
    \rho = \begin{cases}
\ket{\phi}\bra{\phi} &\text{if $k_T=0$}\\
\frac{1}{2}\mathbb{1}_2 &\text{if $k_T = 1$},
\end{cases}
\end{equation}
where $\mathbb{1}_2$ is the identity operator on $\cal{H}_D$ restricted to the subspace spanned by $\left\{ \ket{\mathbb{1}}_D, \ket{S}_D\right\}$. Notice that, were the objects $\cal{T}$ and $\cal{A}$ distinguishable, the device would end up in a maximally mixed state for all combinations of $k_T,k_A$.\\
(c) Finally, Alice measures $\cal{D}$ with projectors $\left\{\Pi_0 = \ket{\phi}\bra{\phi}, \Pi_1= \mathbb{1}_2 - \ket{\phi}\bra{\phi}  \right\}$. If she obtains outcome `0', she guesses $k_T=0$; conversely, if she obtains `1' she guesses $k_T = 1$: the probability of a correct guess is $\frac{3}{4}$, thereby beating a random guess. Can the probability of success be raised closer to unity? The answer is affirmative, and is given by a straightforward extension of the protocol, where $\cal{A}$ now consists of $(N-1)$ additional identical quantum objects, each of them set in the reference state $k_A=0$. The extended protocol is presented in Appendix \ref{appA}, where it is shown that Alice's probability $P_W$ of correctly guessing the required bit is
\begin{equation}
    P_W=1-\frac{1}{2N},
\end{equation}
that asymptotically reaches unity for large $N$. 

Note that if Alice at the end of the process wants to guess which of the objects is the one containing $k_T$ (i.e. $\cal{T}$), she is able to do so only with probability $\frac{1}{N}$. Alternatively, instead of implementing measurement $\left\{\Pi_0, \Pi_1 \right\}$, she could have measured $\cal{D}$ in the $\left\{\ket{\mathbb{1}}_D, \ket{S}_D\right\}$ basis and found out with certainty the location of $k_T$, which would however not have provided her with any knowledge of the value of $k_T$. There is thus a trade-off between the possibility of acquiring knowledge of the value of $k_T$ and of retaining knowledge of its location, which is inherited from the non-commutativity of the observables on $\cal{D}$ that would correspondingly need to be measured.

\subsection*{The necessity of entanglement and indistinguishability}

In the protocol presented in the previous subsection, the interaction between the target objects ($\cal{T}$ and $\cal{A}$) and the device $\cal{D}$ produces a $k_T$-dependent back-reaction on the latter: in particular, the device gets entangled with the target objects if and only if $k_T=1$. This suggests that the possibility of establishing entanglement between the device and the targets, along with the targets being indistinguishable, may be a necessary ingredient for the protocol to work. However, it is not yet clear whether this is the case, since the transformation defined abstractly in Eq. \eqref{eq11} admits other quantum-mechanical realizations besides the ideal quantum control gate that was assumed in Eq. \eqref{eq2}: we thus cannot yet exclude the possibility of there being a noisy control gate that produces a $k_T$-dependent transformation on the device even without the latter getting entangled with the target objects. Nevertheless, as we will sketch here (while leaving the proof for Appendix \ref{appB}), the possibility of establishing entanglement between $\cal{D}$ and the targets turns out after all to be a necessary condition for any quantum-mechanical protocol to solve the task.

Let us assume that $N=2$ and that the two target objects are indistinguishable. As before, states $\ket{\mathbb{1}}_D$ and $\ket{S}_D$ pertaining to $\mathcal{D}$ correspond to the `identity' and `swap' operations. The assumption of our task is that the interaction between $\mathcal{D}$ and the targets is such that, postselecting the final state on $\ket{\mathbb{1}}_D$ leaves the targets' state invariant, whereas postselecting on $\ket{S}_D$ leaves the targets in a `swapped' state. More precisely, the most general allowed interaction $G$ is a CPTP-map that satisfies the following: for any quantum states $\rho^{(D)}$ and $\rho^{(T)}$ of the device $\cal{D}$ and the targets $\cal{T}$ and $\cal{A}$, the post-interaction state $\tilde{\rho} \equiv G(\rho^{(D)} \otimes \rho^{(T)})$ satisfies
\begin{equation}\label{gcontrol}
\begin{split}
    \Tr_D \left[ (\Pi_1 \otimes \mathbb{1}) \tilde{\rho}\right]= p_1 \rho^{(T)}\\
    \Tr_D \left[ (\Pi_S \otimes \mathbb{1}) \tilde{\rho}\right]= p_S \hat{S}\rho^{(T)}\hat{S}^{\dagger},
    \end{split}
\end{equation}
where $\Pi_{1/S}\equiv \ket{\mathbb{1}/S}_D\bra{\mathbb{1}/S}$, and $\hat{S}$ is the `swap' operator acting on the targets. The factors $p_1\equiv \Tr \left(\Pi_1 \rho^{(D)} \right)$ and $p_S\equiv \Tr \left(\Pi_S \rho^{(D)} \right)$ are determined by the additional assumption that the probabilitiy of finding $\cal{D}$ in state $\ket{\mathbb{1}}_D$ or $\ket{S}_D$ does not change upon interaction with the target objects.

Let us label with $\rho^{(T)}_{k_T}$ the initial state of the targets when the unknown bit's value is $k_T$. For any gate $G$ that satisfies Eq. \eqref{gcontrol} and any initial state $\rho^{(D)}$ of $\mathcal{D}$, our task can be solved with probability higher than $\frac{1}{2}$ only if the final reduced state of $\mathcal{D}$ depends on $k_T$, i.e. 
\begin{equation}\label{gcontrol2}
    \Tr_T \left( \tilde{\rho}_0 \right) \neq \Tr_T \left(\tilde{\rho}_1\right),
\end{equation}
where $\tilde{\rho}_{k_T} \equiv G(\rho^{(D)} \otimes \rho^{(T)}_{k_T})$. In Appendix \ref{appB} we show that - for general $N\geq 2$ - while the permutation invariance of state $\rho^{(T)}_0$ implies that $\tilde{\rho}_0$ is a separable state, Eq. \eqref{gcontrol2} holds if and only if $\tilde{\rho}_1$ is an entangled state. Therefore, the necessary conditions for a quantum-mechanical protocol to outperform a random guess are that (i) \textit{it involves indistinguishable target objects}, and that (ii) \textit{the final quantum state of the device and the target objects is entangled if and only if $k_T=1$.}

\section{Discussion}\label{sec3}
We have seen in the previous section that Alice can accomplish her task thanks to the indistinguishability of the target objects and the possibility of establishing entanglement between the latter and the device. We now want to emphasize that the task and its solution are reminiscent of the well known \textit{swap test} \cite{buhrman2001quantum}, which enables one to check whether two quantum systems are prepared in equal states by performing a control-swap operation on them, where the swap operation - henceforth referred to as `abstract swap' - acts as $\ket{\psi}\otimes \ket{\phi} \rightarrow \ket{\phi}\otimes \ket{\psi}$. This offers another angle on how to understand the necessity of the targets' indistinguishability for the quantum protocol to solve our task. Namely, when applied to independently prepared indistinguishable particles, spatially swapping the particles precisely implements, on the space of the particles' spatial modes, the `abstract swap' operation that is required for the swap-test; on the other hand, a spatial swap of independently prepared distinguishable particles strictly differs from an `abstract swap' operation acting on their joint Hilbert space. Stated succinctly, \textit{a spatial swap of two independently prepared particles implements the `abstract swap' operation (of the kind needed in the swap-test) if and only if the particles are indistinguishable}: particle indistinguishability is what enables the operations allowed in our task to be used to implement a standard swap-test. 

As our task presents an instance of quantum indistinguishability serving as an information-theoretic resource, this leads us back to the ongoing debate on the possible physical merit of the entanglement apparently intrinsic to indistinguishable particles, as mentioned in the introductory section. Indeed, notice that our task cannot be solved with distinguishable particles under the requirement that the target objects $\cal{T}$ and $\cal{A}$ be prepared independently. However, if we drop this assumption, the task can be equally well solved with mutually entangled distinguishable particles, as briefly explained in what follows. 

In fact, notice that in the first-quantization formalism, the quantum state associated to a pair of independently prepared indistinguishable particles is mathematically equivalent to a maximally entangled state of ordinary distinguishable particles. Therefore, the quantum-mechanical protocol presented in Section \ref{sec2} - originally interpreted as involving indistinguishable particles - can be straightforwardly re-interpreted, without any change in the mathematical expressions, as involving maximally entangled distinguishable particles.\footnote{For example, the input state in our protocol (i.e. the state on the right hand side of Eq. \eqref{eq4}) is represented in the first-quantization formalism as $\frac{1}{\sqrt{2}}\left(\ket{\vec{x}_A,0}_A \ket{\vec{x}_T,k_T}_T +  \ket{\vec{x}_T,k_T}_A \ket{\vec{x}_A,0}_T \right)$. A mathematically equivalent state can also be assigned to a maximally entangled pair of distinguishable particles.} Indistinguishability and entanglement thus represent equivalent information-theoretic resources in our task, which we take to support the claim that the entanglement that appears in the first quantization notation is more than a mere representational artefact. However, we do not want to delve here into a more detailed discussion on whether independently prepared indistinguishable particles need to (or can) be considered as entangled or not, the answer of which would starkly depend on particular definitions and measures of entanglement \cite{johann2021locality}: our aim is only to point out that \textit{there exists a simple information-theoretic task in which the indistinguishability of independently prepared particles and the entanglement of non-independently prepared distinguishable particles serve as equivalent resources. Importantly, the particles involved in the protocol do not need to spatially overlap and can in fact be arbitrarily distant from each other throughout the whole duration of the protocol}.

Let us now comment on possible future developments of our results. In the current manuscript we have provided only a quantum-mechanical analysis of our task; however, we believe that the latter, as presented abstractly at the beginning of Section \ref{sec2}, can be transposed into the framework of generalized probabilistic theories \cite{barrett2007information, plavala2021general}, which would enable the assessment of how much of our results hinge on the specificities of quantum theory. We may thereby gain a better understanding of the general relationship between indistinguishability and entanglement, by analyzing questions such as the following. In which other operational theories are the indistinguishability of the targets and the possibility of entangling the latter and the device, both necessary to solve the task? In which theories do indistinguishability and entanglement of the targets constitute equivalent resources for our task?

Finally, moving on to potential practical aspects of our findings, it is commonly expected that, following Moore's law, hardware components used in information processing may soon reach a regime in which quantum-mechanical effects cannot be neglected \cite{powell2008quantum}. When (and if) that becomes the case, our results may prove relevant for hardware-security modules that incorporate physical protection against tampering. Examples of this type of devices can already be found both in classical and quantum computing, where the security of one-time programs essentially relies on hardware components, i.e. on \textit{one-time memories} \cite{broadbent2013quantum, roehsner2018quantum}. Other potential applications of our results may be expected to be found in cryptography and hardware security in general.\\

\begin{acknowledgments}
We thank two anonymous referees for insightful comments and for pointing out several useful references. This research was funded in whole, or in part, by the Austrian Science Fund (FWF) ([F7115] and [P36994]). For the purpose of open access, the authors have applied a CC BY public copyright licence to any Author Accepted Manuscript version arising from this submission.
\end{acknowledgments}

\appendix

\begin{widetext}

\section{}\label{appA}
Suppose that Alice is given the object $\mathcal{T}$ in state $\ket{k_T}_{\vec{x}_T}$ and wants to find out the value $k_T$ by using her device $\mathcal{D}$. Let us assume that $\mathcal{A}$ now consists of $(N-1)$ additional identical quantum objects, each of them set in the reference state $k_A=0$ (which is known to Alice), and located at positions $\vec{x_2}$,...,$\vec{x}_{N}$. The joint state of $\mathcal{T}$ together with the $(N-1)$ additional objects is thus $\ket{k_T}_{\vec{x}_1}\ket{0}_{\vec{x}_2}...\ket{0}_{\vec{x}_{N}}$, where we have for simplicity renamed position $\vec{x}_T$ into $\vec{x}_1$. Alice's device $\mathcal{D}$ can now be used to swap any pair of the target objects: let us label with `$S_i$' the state of $\mathcal{D}$ that is set to swap $\mathcal{T}$ with the $i$-th object. The protocol is then partitioned in three steps:\\
(a) Alice prepares $\mathcal{D}$ in state $\ket{\phi} = \sum_{i=1}^{N} \ket{S_i}_D$, where $\ket{S_1}_D \equiv \ket{\mathbb{1}}_D$.\\
(b) She lets $\mathcal{D}$, $\mathcal{T}$ and the $(N-1)$ additional objects interact as:
\begin{equation}\label{eq8}
\frac{1}{\sqrt{N}}\sum_{i=1}^{N} \ket{S_i}_D \ket{k_T}_{\vec{x}_1} \ket{0}_{\vec{x}_2}...\ket{0}_{\vec{x}_{N}} \rightarrow \frac{1}{\sqrt{N}}\sum_{i=1}^{N} \ket{S_i}_D \ket{0}_{\vec{x}_1} \ket{0}_{\vec{x}_2}...\ket{k_T}_{\vec{x}_i}...\ket{0}_{\vec{x}_{N}}.
\end{equation}
After the interaction, the reduced density state $\rho$ associated to $\mathcal{D}$ is:
\begin{equation}\label{eq9}
    \rho = \begin{cases}
\ket{\phi}\bra{\phi} &\text{if $k_T=0$}\\
\frac{1}{N}\mathbb{1}_{N} &\text{if $k_T \neq 0$},
\end{cases}
\end{equation}
where $\mathbb{1}_{N}$ is the identity operator on $\mathcal{H}_D$ when restricted to the corresponding subspace.\\
(c) Finally, Alice measures $\mathcal{D}$ with projectors $\left\{\Pi_0 = \ket{\phi}\bra{\phi}, \Pi_1= \mathbb{1}_D - \ket{\phi}\bra{\phi}  \right\}$. If she obtains outcome `0', she guesses $k_T=0$; conversely, if she obtains `1' she guesses $k_T = 1$. The probability of a correct guess $P_W$ is 
\begin{equation}
    P_W=1-\frac{1}{2N},
\end{equation}
which coincides with the maximum possible value given by the Helstrom bound \cite{helstrom1969quantum}, thereby showing that the above measurement is optimal.

\section{}\label{appB}
Here we will provide the proof that in all quantum-mechanical protocols that solve our task with probability higher than $\frac{1}{2}$, the device and the targets get entangled upon interaction in the case that $k=1$. We will delve immediately into the general case, where $\mathcal{A}$ consists of $(N-1)$ identical quantum systems set in state $0$ and located at positions $\vec{x}_2$,...,$\vec{x}_{N}$, whereas $\mathcal{T}$ is set in state $k$ and located at position $\vec{x}_1$. The joint initial state of the targets is thus $\ket{k}_{\vec{x}_1} \ket{0}_{\vec{x}_2}...\ket{0}_{\vec{x}_{N}}$. Let us for simplicity introduce the following notation:
\begin{equation}
\begin{split}\label{a1}
    &\ket{0}_T\equiv \ket{0}_{\vec{x}_1}\ket{0}_{\vec{x}_2} ...\ket{0}_{\vec{x}_{N}}\\
    &\ket{i}_T\equiv \ket{0}_{\vec{x}_1}\ket{0}_{\vec{x}_2} ...\ket{1}_{\vec{x}_i} ... \ket{0}_{\vec{x}_{N}}, \quad i=1,...,N.
    \end{split}
\end{equation}
Since by assumption Alice is only able to coherently swap the targets, the effective Hilbert space $\mathcal{H}_T$ that can be associated to the latter is spanned by the vectors in Eq. \eqref{a1}, i.e. $\mathcal{H}_T=\text{Span}(\ket{i}_T, i=0,...,N)$. A swap gate $S_j$ that swaps the first and $j$-th object thus acts as $S_j \ket{0}_T=\ket{0}_T$, and $S_j \ket{1}_T=\ket{j}_T$ (notice that $S_1$ is the identity operator). The initial targets' state is equal to $\ket{k}_T$, with $k$ being 0 or 1. Finally, the Hilbert space $\mathcal{H}_D$ associated to the device $\mathcal{D}$ is spanned by vectors $\left\{\ket{j}_D, j=1,...,N \right\}$, with each $\ket{j}_D$ representing the setting corresponding to swap gate $S_j$.

Now we want to characterize the most general interaction between $\mathcal{D}$ and the targets that is allowed by the assumptions of our task. Let $\mathbf{St}_{DT}$, $\mathbf{St}_{D}$ and $\mathbf{St}_{T}$ be respectively sets of density operators on $\mathcal{H}_D \otimes \mathcal{H}_T$, $\mathcal{H}_D$ and $\mathcal{H}_T$. We will label with $\mathbf{Ch}_{DT}$ the set of completely positive trace preserving (CPTP) superoperators that map $\mathbf{St}_{DT}$ into itself. Let $\mathcal{G}$ be the subset of CPTP maps that represent the device-targets interactions implementable by Alice. The abstract formulation of the task implies that, for any $G \in \mathcal{G}$, if the initial state of $\mathcal{D}$ is some $\ket{i}_D$, then the interaction is given by 
\begin{equation}\label{g1}
    G\left(\ket{i}_D\bra{i}\otimes \rho^{(T)}\right)=\ket{i}_D\bra{i}\otimes S_i\rho^{(T)}S_i,
\end{equation}
for any $\rho^{(T)} \in \mathbf{St}_{T}$, where we have used the property $S^{\dagger}_i=S_i$. More generally, if the initial state of $\mathcal{D}$ is an arbitrary density operator $\rho^{(D)} \in \mathbf{St}_{D}$, then the interaction is such, that postselecting the final state on the device's state $\ket{i}_D$ implements gate $S_i$ on the targets, i.e.
\begin{equation}\label{g2}
    \Tr_D \left[ \left(\Pi_i \otimes \mathbb{1} \right) G\left(\rho^{(D)} \otimes \rho^{(T)} \right) \right] = p_i S_i\rho^{(T)}S_i,
\end{equation}
where $\Pi_i\equiv \ket{i}_D\bra{i}$. The factors $p_i= \Tr \left(\Pi_i \rho^{(D)} \right)$ are determined by the further assumption that the probabilitiy of finding $\cal{D}$ in state $\ket{i}_D$ is invariant upon interaction with the targets.

Therefore, the assumptions of the task imply that the set $\mathcal{G}$ of allowed device-targets interactions can be characterized as follows:
\begin{equation}\label{set of gates}
    \mathcal{G}=\left\{G \in  \mathbf{Ch}_{DT} \vert \quad \forall \rho^{(D)} \in \mathbf{St}_{D}, \rho^{(T)} \in \mathbf{St}_{T}, i=1,...,N:\quad \Tr_D \left[ \left(\Pi_i \otimes \mathbb{1} \right) G\left(\rho^{(D)} \otimes \rho^{(T)} \right) \right] = p_i S_i\rho^{(T)}S_i\right\},
\end{equation}
where $p_i\equiv\Tr \left(\Pi_i \rho^{(D)} \right)$. $\mathcal{G}$ can be understood as the class of \textit{generalized control gates}, with $\mathcal{D}$ being the control system and $\mathcal{A}$ and $\mathcal{T}$ constituting the target system. In the main text we have seen that in the case of the ideal control gate, which is a particular element of $\mathcal{G}$, the device and the targets get entangled upon interaction if $k=1$; in what follows we will prove that this is the case for any element of $\mathcal{G}$ that can be used to outperform a random guess at solving our task. Let us first slightly generalize the discussion by defining a broader class of generalized control gates for which we will prove a Lemma that will be of use later.\\
\\
\begin{Definition}\label{def}
Consider $\mathcal{H}_D\equiv \mathbb{C}^{N_D}$ and $\mathcal{H}_T\equiv \mathbb{C}^{N_T}$ for some finite natural numbers $N_D$ and $N_T$. Let $\mathbf{St}_{DT}$, $\mathbf{St}_{D}$ and $\mathbf{St}_{T}$ be sets containing all and only density operators on respectively $\mathcal{H}_D \otimes \mathcal{H}_T$, $\mathcal{H}_D$ and $\mathcal{H}_T$. Let $\mathbf{Ch}_{DT}$ be the set of completely positive trace preserving (CPTP) superoperators that map $\mathbf{St}_{DT}$ into itself.\\ 
Let $\vec{U}\equiv \langle U^{(1)},...,U^{(N_D)} \rangle$ be a list of $N_D$ unitary operators on $\mathcal{H}_T$, and let $\vec{\Pi}\equiv \langle\Pi_1,...,\Pi_{N_D}\rangle$ be a list of projectors that form a projection-valued measure on $\mathcal{H}_D$. Define a 4-tuple $X$ as $X\equiv \langle N_D,N_T,\vec{U},\vec{\Pi}\rangle$.\\
The \textbf{set of generalized control gates relative to $X$} is a set $\mathcal{G}_X$ defined as
\begin{equation*}
\mathcal{G}_X=\left\{G \in  \mathbf{Ch}_{DT} \vert \quad \forall \rho^{(D)} \in \mathbf{St}_{D}, \rho^{(T)} \in \mathbf{St}_{T}, i=1,...,N_D:\quad \Tr_D \left[ \left(\Pi_i \otimes \mathbb{1} \right) G\left(\rho^{(D)} \otimes \rho^{(T)} \right) \right] = p_i U^{(i)}\rho^{(T)}U^{(i)\dagger}\right\},
\end{equation*}
where $p_i=\Tr \left(\Pi_i \rho^{(D)} \right)$.\\
\end{Definition}

\begin{Lemma}\label{lem}
Consider a 4-tuple $X\equiv \langle N_D,N_T,\vec{U},\vec{\Pi}\rangle$, where $N_D,N_T$ are natural numbers, $\vec{U}\equiv \langle U^{(1)},...,U^{(N_D)} \rangle$ is a list of unitary operators on $\mathcal{H}_T\equiv \mathbb{C}^{N_T}$, and $\vec{\Pi}\equiv \langle \Pi_1,...,\Pi_{N_D} \rangle$ is a projection-valued measure on $\mathcal{H}_D\equiv \mathbb{C}^{N_D}$. Let $\mathcal{G}_X$ be the set of generalized control gates relative to $X$.\\ 
Then, for any $G\in \mathcal{G}_X$, there exists a natural number $M$ and a set $\left\{\vec{v}^{(1)},...,\vec{v}^{(N_D)}\right\}$ of unit vectors in $\mathbb{C}^{M}$, such that, for all $\rho^{(D)} \in \mathbf{St}_{D}$ and $\rho^{(T)} \in \mathbf{St}_{T}$:
\begin{equation*}
    G\left(\rho^{(D)} \otimes \rho^{(T)} \right)= \sum_{k,l=1}^{N_D} \vec{v}^{(k)}\cdot \vec{v}^{(l)} \rho^{(D)}_{kl} \ket{k}\bra{l} \otimes U^{(k)} \rho^{(T)} U^{(l)\dagger},
\end{equation*}
where $\vec{v}^{(k)}\cdot \vec{v}^{(l)}=\sum_{j=1}^{M} v_j^{(k)} v_j^{(l)*}$ is the dot product in $\mathbb{C}^M$, and $\rho^{(D)}_{kl}$ are the components of $\rho^{(D)}$ in the basis $\left\{ \ket{1},...,\ket{N_D} \right\}$ associated to the list of projectors $\vec{\Pi}$.
\end{Lemma}

\bigskip

\begin{proof}
Take arbitrary elements $G \in \mathcal{G}_X$, $\rho^{(D)} \in \mathbf{St}_{D}$ and $\rho^{(T)} \in \mathbf{St}_{T}$. Since $G$ is a CPTP operator, then, for some $M \geq 1$, there exists a set of Kraus operators $\left\{K_1,...,K_M \right\}$ on $\mathcal{H}_D \otimes \mathcal{H}_T$ that satisfy $\sum_j K_j^{\dagger}K_j = \mathbb{1}$ and
\begin{equation}\label{p1}
    G\left(\rho^{(D)} \otimes \rho^{(T)} \right)=\sum_{j=1}^{M} K_j \left(\rho^{(D)} \otimes \rho^{(T)} \right) K_j^{\dagger}.
\end{equation}
For each Kraus operator $K_j$ let us introduce a set of $(N_D)^2$ operators $\left\{B_j^{(kl)}\right\}$ on $\mathcal{H}_T$, such that the following holds:
\begin{equation}\label{p2}
    K_j=\sum_{k,l=1}^{N_D} \ket{k}\bra{l} \otimes B_j^{(kl)},
\end{equation}
where the basis vectors $\left\{\ket{1},...,\ket{N_D}\right\}$ are the ones associated to the projectors $\vec{\Pi}$. Plugging in decomposition \eqref{p2} into the Kraus representation \eqref{p1} we obtain
\begin{equation}\label{p3}
    G\left(\rho^{(D)} \otimes \rho^{(T)} \right)=\sum_{j=1}^{M}\sum_{k,l,n,m=1}^{N_D} \rho_{lm}^{(D)} \ket{k}\bra{l} \otimes B_j^{(kl)} \rho^{(T)} B_j^{(nm)\dagger}.
\end{equation}
Since $G$ is an element of $\mathcal{G}$, it satisfies 
\begin{equation}\label{p4}
    \Tr_D \left[ \left(\Pi_s \otimes \mathbb{1} \right) G\left(\rho^{(D)} \otimes \rho^{(T)} \right) \right] = \Tr_D \left(\Pi_s \rho^{(D)} \right) U^{(s)}\rho^{(T)}U^{(s)\dagger},
\end{equation}
for any $s=1,...,N_D$. Together with Eq. \eqref{p3}, this implies that the following holds for all $\rho^{(D)}, \rho^{(T)}$:
\begin{equation}\label{p5}
    \sum_{j=1}^{M}\sum_{l,m=1}^{N_D} \rho_{lm}^{(D)} \otimes B_j^{(sl)} \rho^{(T)} B_j^{(sm)\dagger}=\rho_{ss}^{(D)}U^{(s)}\rho^{(T)}U^{(s)\dagger}.
\end{equation}
Since the latter is valid for all density operators $\rho^{(D)}$, then it must also hold for for all operators in $\left\{ \ket{a}\bra{b}, a,b=1...,N_D \right\}$, as the latter constitutes a basis for the space of density operators on $\mathcal{H}_D$. Therefore, for all $s,a,b=1...,N_D$, and all $\rho^{(T)}$:
\begin{equation}\label{p6}
    \sum_{j=1}^{M} B_j^{(sa)} \rho^{(T)} B_j^{(sb)\dagger} = \delta_{s,a}\delta_{s,b} U^{(s)}\rho^{(T)} U^{(s)\dagger}.
\end{equation}
It can be easily seen that Eq. \eqref{p6} implies that, for all $j$ and $s\neq a$: 
\begin{equation}\label{p7}
    B_j^{(sa)}=0,
\end{equation}
which motivates us to define a new set of operators $\left\{B^{(s)}_j \right\}$ on $\mathcal{H}_T$, such that $B_j^{(sa)}=\delta_{s,a}B_j^{(s)}$. Eq. \eqref{p6} then implies that for all $s=1,...,N_D$:
\begin{equation}\label{p8}
    \sum_{j=1}^{M} B_j^{(s)} \rho^{(T)}B_j^{(s)\dagger}=U^{(s)} \rho^{(T)} U^{(s)\dagger}.
\end{equation}
Getting back to the initially introduced Kraus operators, it is simple to deduce from the normalization condition $\sum_j K_j^{\dagger}K_j = \mathbb{1}$ that the auxiliary operators $B_j^{(kl)}$ satisfy 
\begin{equation}\label{p9}
    \sum_{j=1}^{M}\sum_{k=1}^{N_D} B_j^{(kl)\dagger} B_j^{(kn)}=\delta_{l,n} \mathbb{1},
\end{equation}
for all $l,n=1,...,N_D$. This in turn implies that the newly defined operators $\left\{B^{(s)}_j \right\}$ satisfy
\begin{equation}\label{p10}
    \sum_{j=1}^{M} B_j^{(s)\dagger}B_j^{(s)}=\mathbb{1}.
\end{equation}
Eqs. \eqref{p8} and \eqref{p10} entail that for each $s$, the set $\left\{B_j^{(s)}, j=1,...,M \right\}$ is a Kraus representation of the CPTP map $\mathcal{C}^{(s)}:\textbf{St}_T \rightarrow \textbf{St}_T$ that acts as
\begin{equation}\label{p11}
    \mathcal{C}^{(s)}\left( \rho^{(T)}\right)=U^{(s)} \rho^{(T)} U^{(s)\dagger},
\end{equation}
for all $\rho^{(T)} \in \textbf{St}_T$. One Kraus representation of $\mathcal{C}^{(s)}$ is trivially given by $B_j^{(s)}=\delta_{j,1}U^{(s)}$. Now recall that all Kraus representations of a CPTP map are unitarily equivalent, meaning that for any two representations $\left\{B_j\right\}$ and $\left\{A_j\right\}$ of the same map, there exists a unitary operator $u$ on $\mathbb{C}^M$, such that $A_i=\sum_{j=1}^{M}u_{ij}B_j$. Therefore, for all $s$, all sets that satisfy Eqs. \eqref{p8} and \eqref{p10} are unitarily equivalent to $\left\{\delta_{j,1}U^{(s)}\right\}$, which means that a set $\left\{B_j^{(s)}, j=1,...,M \right\}$ satisfies Eqs. \eqref{p8} and \eqref{p10} if and only if there exists a unitary operator $u^{(s)}$, such that $B_j^{(s)}=u^{(s)}_{j1}U^{(s)}$. Since the column $\left(u^{(s)}_{11},...,u^{(s)}_{M1}\right)$ of any unitary operator $u^{(s)}$ on $\mathbb{C}^{M}$ is just a unit vector in $\mathbb{C}^{M}$, it follows that $\left\{B_j^{(s)} \right\}$ is a viable solution if and only if there exists a unit vector $\vec{v} \in \mathbb{C}^{M}$ such that $B_j^{(s)}=v^{(s)}_{j}U^{(s)}$, with $v^{(s)}_{j}$ being the $j$-th component of vector $\vec{v}^{(s)}$ in some basis.\\
It thus follows from the defining Eq. \eqref{p2} that for any $G \in \mathcal{G}_X$, and any $M$-element-set of Kraus operators $\left\{K_1,...,K_M\right\}$ of $G$, there exists a set $\left\{\vec{v}_1,...,\vec{v}_{N_D}\right\}$ of unit vectors in $\mathbb{C}^M$, such that
\begin{equation}\label{p12}
    K_j=\sum_{s=1}^{N_D} v^{(s)}_j \ket{s}\bra{s} \otimes U^{(s)},
\end{equation}
for all $j=1,...,M$. Eq. \eqref{p12}, together with the definition of the Kraus representation, entails that for all $G \in \mathcal{G}$, there exists a natural number $M \geq 1$ and a set $\left\{\vec{v}_1,...,\vec{v}_{N_D}\right\}$ of unit vectors in $\mathbb{C}^M$, such that  for all $\rho^{(D)} \in \mathbf{St}_D$ and $\rho^{(T)} \in \mathbf{St}_T$:
\begin{equation}\label{p13}
    G\left(\rho^{(D)} \otimes \rho^{(T)} \right)= \sum_{k,l=1}^{N_D} \vec{v}^{(k)}\cdot \vec{v}^{(l)} \rho^{(D)}_{kl} \ket{k}\bra{l} \otimes U^{(k)} \rho^{(T)} U^{(l)\dagger},
\end{equation}
where $\vec{v}^{(k)}\cdot \vec{v}^{(l)}=\sum_{j=1}^{M} v_j^{(k)} v_j^{(l)*}$ is the dot product in $\mathbb{C}^M$.
\end{proof}

\bigskip
\bigskip
Let us briefly comment on two special cases of Eq. \eqref{p13}. Consider a transformation $G$, which is such that all its corresponding vectors $\vec{v}^{(i)}$ are equal, i.e. $\vec{v}^{(i)} \cdot \vec{v}^{(j)}=1$ for all $i,j$. Then it is easy to see that $G$ represents an ideal unitary gate, such as the one that we have used in the main text's protocol, i.e.
\begin{equation}
    G(\rho^{(D)} \otimes \rho^{(T)})=U(\rho^{(D)} \otimes \rho^{(T)})U^{\dagger}, \quad U=\sum_{k=1}^{N_D}\ket{k}\bra{k} \otimes U^{(k)}.
\end{equation}
On the other hand, consider a gate $G$ whose vectors are all mutually orthogonal, i.e. $\vec{v}^{(i)} \cdot \vec{v}^{(j)}=\delta_{i,j}$; then $G$ acts as
\begin{equation}
    G(\rho^{(D)} \otimes \rho^{(T)})=\sum_{k=1}^{N_D} \rho^{(D)}_{kk} \ket{k}\bra{k} \otimes U^{(k)} \rho^{(T)} U^{(k)\dagger}.
\end{equation}
The latter gate erases the off-diagonal terms of $\rho^{(D)}$ in the basis defined by projectors $\vec{\Pi}$. In other words, it decoheres the device in the aforementioned basis, and may thus as well be understood as a ``classical control gate''.\\
\\
Let us now use the obtained results to prove that any protocol that solves our task necessarily establishes entanglement between the target and the devices if $k=1$. Applying Definition \ref{def} to Eq. \eqref{set of gates}, we see that the set $\mathcal{G}_Y \equiv \mathcal{G}$ of generalized control gates that are allowed in our task is defined relative to the 4-tuple $Y\equiv \langle N_D,N_T,\vec{U},\vec{\Pi}\rangle$, with the following identifications: $N_D=N$ and $N_T=N+1$, where $N$ is the number of target objects; $\vec{U}=\langle S_1,...,S_N \rangle$ where the unitary operators $S_i$ are swap-gates; and $\vec{\Pi}=\langle \Pi_1,...,\Pi_N \rangle$ is the set of projectors on states representing settings associated to the corresponding swap-gates. Lemma \ref{lem} then implies that for any $G \in \mathcal{G}_Y$, there exists a natural number $M \geq 1$ and a set $\left\{\vec{v}^{(1)},...,\vec{v}^{(N_D)}\right\}$ of unit vectors in $\mathbb{C}^M$, such that  for all $\rho^{(D)} \in \mathbf{St}_D$ and $\rho^{(T)} \in \mathbf{St}_T$:
\begin{equation}\label{z1}
    G\left(\rho^{(D)} \otimes \rho^{(T)} \right)= \sum_{k,l=1}^{N} \vec{v}^{(k)}\cdot \vec{v}^{(l)} \rho^{(D)}_{kl} \ket{k}\bra{l} \otimes S_k \rho^{(T)} S_l,
\end{equation}
where $\vec{v}^{(k)}\cdot \vec{v}^{(l)}=\sum_{j=1}^{M} v_j^{(k)} v_j^{(l)*}$ is the dot product in $\mathbb{C}^M$. If the unknown bit $k$ is equal to 0, then $\rho^{(T)}$ is equal to $\rho^{(T)}_0=\ket{0}\bra{0}$. Therefore, since $S_k \ket{0}\bra{0} S_l=\ket{0}\bra{0}$, then
\begin{equation}\label{z2}
    G\left(\rho^{(D)} \otimes \rho^{(T)}_0 \right)= \sum_{k,l=1}^{N} \vec{v}^{(k)}\cdot \vec{v}^{(l)} \rho^{(D)}_{kl} \ket{k}\bra{l} \otimes \rho^{(T)}_0.
\end{equation}
On the other hand, if $k=1$, then $\rho^{(T)}$ is equal to $\rho^{(T)}_1=\ket{1}\bra{1}$. Since $S_k \ket{1}\bra{1} S_l=\ket{k}\bra{l}$, then 
\begin{equation}\label{z3}
    G\left(\rho^{(D)} \otimes \rho^{(T)}_1 \right)= \sum_{k,l=1}^{N} \vec{v}^{(k)}\cdot \vec{v}^{(l)} \rho^{(D)}_{kl} \ket{k}\bra{l} \otimes \ket{k}\bra{l}.
\end{equation}
If we define the corresponding reduced density states of $\mathcal{D}$ as $\tilde{\rho}^{(D)}_k\equiv \Tr_T \left[G\left(\rho^{(D)} \otimes \rho^{(T)}_k \right)\right]$, then Eqs. \eqref{z2} and \eqref{z3} imply
\begin{equation}\label{z4}
\begin{split}
    & \tilde{\rho}^{(D)}_0=\sum_{i,j=1}^{N}\alpha_{ij}\ket{i}\bra{j},\\
    & \tilde{\rho}^{(D)}_1=\sum_{i}^{N}\alpha_{ii}\ket{i}\bra{j},
\end{split}
\end{equation}
where $\alpha_{ij}\equiv \vec{v}^{(i)} \cdot \vec{v}^{(j)} \rho^{(D)}_{ij}$.\\
In order to be able to acquire some information about the value $k$ via a measurement on $\mathcal{D}$, the two states $\tilde{\rho}^{(D)}_0$ and $\tilde{\rho}^{(D)}_1$ cannot be equal. Therefore, Eq. \eqref{z4} implies that the probability $P_W$ of successfully solving the task can be higher than $\frac{1}{2}$ if and only if there exists at least one pair $n\neq m$, such that $\alpha_{nm}\neq 0$. The latter is equivalent to the requirement that $\rho^{(D)}_{nm}\neq 0$ and $\vec{v}^{(n)} \cdot \vec{v}^{(m)} \neq 0$, which means that the initial state of $\mathcal{D}$ needs to have some coherence in the preferred basis chosen by projectors $\vec{\Pi}$, and that the gate cannot fully decohere the control in the latter basis, i.e. it cannot be a ``classical control gate''. This establishes a sense in which the task cannot be solved classically: the initial state of the control needs to be ``non-classical'' in the sense of having coherence in the aforementioned basis, and the interaction between the device and the targets need to be non-classical, in the sense of preserving the device's coherence terms.\\
Now we are finally ready to inspect whether it is possible to gain some information about $k$ without establishing entanglement between the device and the targets when $k=1$. In other words, if $\alpha_{nm}\neq 0$ for at least one pair $n\neq m$, is the state $G\left(\rho^{(D)} \otimes \rho^{(T)}_1 \right)$ necessarily entangled? We are going to answer this question by employing the PPT criterion, which states that a necessary condition for a quantum state $\rho$ defined on $\mathcal{H}_A \otimes \mathcal{H}_B$ to be separable is that the partially transposed state $\rho^{T_B}\equiv (\mathbb{1}\otimes T)\rho$ does not have negative eigenvalues, where the transpose $T$ acts as $T\ket{i}\bra{j}=\ket{j}\bra{i}$ \cite{peres1996separability}. Consequently, a sufficient (but generally not necessary) condition for a state to be entangled is that its partial transpose has at least one negative eigenvalue.\\
Let us apply the partial transpose operation on $\rho_1\equiv G\left(\rho^{(D)} \otimes \rho^{(T)}_1 \right)$, thereby obtaining
\begin{equation}\label{final}
    (\mathbb{1}\otimes T)\rho_1=\sum_j \alpha_{jj} \ket{j}\bra{j} \otimes \ket{j}\bra{j}+\sum_{m<n} \alpha_{mn} \ket{m}\bra{n} \otimes \ket{n}\bra{m} + \sum_{m<n} \alpha_{mn}^* \ket{n}\bra{m} \otimes \ket{m}\bra{n}.
\end{equation}
Now suppose that there exist $k\neq l$, for which $\alpha_{kl}\neq 0$. Then it is simple to check that the vector $\ket{\psi}_{kl}$ defined as
\begin{equation}
    \ket{\psi}_{kl}\equiv \ket{k}\otimes \ket{l} - \frac{\alpha_{kl}^*}{|\alpha_{kl}|}\ket{l}\otimes \ket{k}
\end{equation}
is an eigenvector of $(\mathbb{1}\otimes T)\rho_1$ with its corresponding eigenvalue equal to $-|\alpha_{kl}|$, which is negative. Therefore, if $\alpha_{kl}\neq 0$ for some $k \neq l$, then the state $\rho_1$ is necessarily entangled. We thereby completed the proof that a quantum-mechanical protocol that involves indistinguishable particles can outperform a random guess if and only if the device and the targets get entangled upon interaction when $k=1$.
\\

\end{widetext}

\begin{thebibliography}{10}

\bibitem {her2001} Herbut, F., 2001. How to distinguish identical particles. American Journal of Physics, 69(2), pp.207-217.

\bibitem {her2006} Herbut, F., 2006. How to distinguish identical particles. The general case. arXiv preprint quant-ph/0611049.

\bibitem {ghirardi2002entanglement} Ghirardi, G., Marinatto, L. and Weber, T., 2002. Entanglement and properties of composite quantum systems: a conceptual and mathematical analysis. Journal of statistical Physics, 108(1), pp.49-122.

\bibitem {ghir2003} Ghirardi, G. and Marinatto, L., 2003. Entanglement and properties. Fortschritte der Physik: Progress of Physics, 51(4‐5), pp.379-387.

\bibitem {ghirardi2004general} Ghirardi, G. and Marinatto, L., 2004. General criterion for the entanglement of two indistinguishable particles. Physical Review A, 70(1), p.012109.

\bibitem {johann2021locality} Johann, T.J.F. and Marzolino, U., 2021. Locality and entanglement of indistinguishable particles. Scientific Reports, 11(1), pp.1-10.

\bibitem {benatti2021entanglement} Benatti, F., Floreanini, R. and Marzolino, U., 2021. Entanglement and non-locality in quantum protocols with identical particles. Entropy, 23(4), p.479.

\bibitem {benatti2020entanglement} Benatti, F., Floreanini, R., Franchini, F. and Marzolino, U., 2020. Entanglement in indistinguishable particle systems. Physics Reports, 878, pp.1-27.

\bibitem {li2001} Li, Y.S., Zeng, B., Liu, X.S. and Long, G.L., 2001. Entanglement in a two-identical-particle system. Physical Review A, 64(5), p.054302.

\bibitem {pas2001} Paškauskas, R. and You, L., 2001. Quantum correlations in two-boson wave functions. Physical Review A, 64(4), p.042310.

\bibitem {schl2001} Schliemann, J., Cirac, J.I., Kuś, M., Lewenstein, M. and Loss, D., 2001. Quantum correlations in two-fermion systems. Physical Review A, 64(2), p.022303.

\bibitem {eck2002} Eckert, K., Schliemann, J., Bruß, D. and Lewenstein, M., 2002. Quantum correlations in systems of indistinguishable particles. Annals of physics, 299(1), pp.88-127.

\bibitem {gitt2002} Gittings, J.R. and Fisher, A.J., 2002. Describing mixed spin-space entanglement of pure states of indistinguishable particles using an occupation-number basis. Physical Review A, 66(3), p.032305.

\bibitem {zan2002} Zanardi, P., 2002. Quantum entanglement in fermionic lattices. Physical review A, 65(4), p.042101.

\bibitem {omar2002} Omar, Y., Paunković, N., Bose, S. and Vedral, V., 2002. Spin-space entanglement transfer and quantum statistics. Physical Review A, 65(6), p.062305.

\bibitem {paun2002} Paunković, N., Omar, Y., Bose, S. and Vedral, V., 2002. Entanglement concentration using quantum statistics. Physical review letters, 88(18), p.187903.

\bibitem {shi2003} Shi, Y., 2003. Quantum entanglement of identical particles. Physical Review A, 67(2), p.024301.

\bibitem {ved2003} Vedral, V., 2003. Entanglement in the second quantization formalism. Central European Journal of Physics, 1, pp.289-306.

\bibitem {wise2003} Wiseman, H.M. and Vaccaro, J.A., 2003. Entanglement of indistinguishable particles shared between two parties. Physical review letters, 91(9), p.097902.

\bibitem {bose2003} Bose, S., Ekert, A., Omar, Y., Paunković, N. and Vedral, V., 2003. Optimal state discrimination using particle statistics. Physical Review A, 68(5), p.052309.

\bibitem {sher2006} Sheridan, L., Paunković, N., Omar, Y. and Bose, S., 2006. Discrete time quantum walk on a line with two particles. International Journal of Quantum Information, 4(03), pp.573-583.

\bibitem {omar2006} Omar, Y., Paunković, N., Sheridan, L. and Bose, S., 2006. Quantum walk on a line with two entangled particles. Physical Review A, 74(4), p.042304.

\bibitem {cavalcanti2007useful} Cavalcanti, D., Malard, L.M., Matinaga, F.M., Cunha, M.T. and Santos, M.F., 2007. Useful entanglement from the Pauli principle. Physical Review B, 76(11), p.113304.

\bibitem {tichy2013} Tichy, M.C., de Melo, F., Kuś, M., Mintert, F. and Buchleitner, A., 2013. Entanglement of identical particles and the detection process. Fortschritte der Physik, 61(2‐3), pp.225-237.

\bibitem {krenn2017} Krenn, M., Hochrainer, A., Lahiri, M. and Zeilinger, A., 2017. Entanglement by path identity. Physical review letters, 118(8), p.080401.

\bibitem {li2017} Li, X.M., Yang, M., Paunković, N., Li, D.C. and Cao, Z.L., 2017. Entanglement swapping via three-step quantum walk-like protocol. Physics Letters A, 381(46), pp.3875-3879.

\bibitem {franco2018indistinguishability} Franco, R.L. and Compagno, G., 2018. Indistinguishability of elementary systems as a resource for quantum information processing. Physical review letters, 120(24), p.240403.

\bibitem {chin2019} Chin, S. and Huh, J., 2019. Entanglement of identical particles and coherence in the first quantization language. Physical Review A, 99(5), p.052345.

\bibitem {kar2019} Karczewski, M., Lee, S.Y., Ryu, J., Lasmar, Z., Kaszlikowski, D. and Kurzyński, P., 2019. Sculpting out quantum correlations with bosonic subtraction. Physical Review A, 100(3), p.033828.

\bibitem {ju2019} Ju, L., Yang, M., Paunković, N., Chu, W.J. and Cao, Z.L., 2019. Creating photonic GHZ and W states via quantum walk. Quantum Information Processing, 18, pp.1-12.

\bibitem {barghathi2019operationally} Barghathi, H., Casiano-Diaz, E. and Del Maestro, A., 2019. Operationally accessible entanglement of one-dimensional spinless fermions. Physical Review A, 100(2), p.022324.

\bibitem {castellini2019} Castellini, A., Bellomo, B., Compagno, G. and Franco, R.L., 2019. Activating remote entanglement in a quantum network by local counting of identical particles. Physical Review A, 99(6), p.062322.

\bibitem {castellini2019b} Castellini, A., Franco, R.L., Lami, L., Winter, A., Adesso, G. and Compagno, G., 2019. Indistinguishability-enabled coherence for quantum metrology. Physical Review A, 100(1), p.012308.

\bibitem {nosrati2020a} Nosrati, F., Castellini, A., Compagno, G. and Lo Franco, R., 2020. Robust entanglement preparation against noise by controlling spatial indistinguishability. npj Quantum Information, 6(1), p.39.

\bibitem {nosrati2020b} Nosrati, F., Castellini, A., Compagno, G. and Franco, R.L., 2020. Dynamics of spatially indistinguishable particles and quantum entanglement protection. Physical Review A, 102(6), p.062429.

\bibitem {morris2020entanglement} Morris, B., Yadin, B., Fadel, M., Zibold, T., Treutlein, P. and Adesso, G., 2020. Entanglement between identical particles is a useful and consistent resource. Physical Review X, 10(4), p.041012.

\bibitem {barros2020} Barros, M.R., Chin, S., Pramanik, T., Lim, H.T., Cho, Y.W., Huh, J. and Kim, Y.S., 2020. Entangling bosons through particle indistinguishability and spatial overlap. Optics Express, 28(25), pp.38083-38092.

\bibitem {sun2020} Sun, K., Wang, Y., Liu, Z.H., Xu, X.Y., Xu, J.S., Li, C.F., Guo, G.C., Castellini, A., Nosrati, F., Compagno, G. and Franco, R.L., 2020. Experimental quantum entanglement and teleportation by tuning remote spatial indistinguishability of independent photons. Optics Letters, 45(23), pp.6410-6413.

\bibitem {holmes2020} Holmes, Z., Anders, J. and Mintert, F., 2020. Enhanced energy transfer to an optomechanical piston from indistinguishable photons. Physical Review Letters, 124(21), p.210601.

\bibitem {chin2021} Chin, S., Kim, Y.S. and Lee, S., 2021. Graph picture of linear quantum networks and entanglement. Quantum, 5, p.611.

\bibitem {wang2022} Wang, Y., Hao, Z.Y., Liu, Z.H., Sun, K., Xu, J.S., Li, C.F., Guo, G.C., Castellini, A., Bellomo, B., Compagno, G. and Franco, R.L., 2022. Remote entanglement distribution in a quantum network via multinode indistinguishability of photons. Physical Review A, 106(3), p.032609.

\bibitem {zaw2022} Zaw, L.H., Lasmar, Z., Nguyen, C.H., Tseng, K.W., Matsukevich, D., Kaszlikowski, D. and Scarani, V., 2022. Sculpting bosonic states with arithmetic subtractions. New Journal of Physics, 24(8), p.083023.

\bibitem {sun2022} Sun, K., Liu, Z.H., Wang, Y., Hao, Z.Y., Xu, X.Y., Xu, J.S., Li, C.F., Guo, G.C., Castellini, A., Lami, L. and Winter, A., 2022. Activation of indistinguishability-based quantum coherence for enhanced metrological applications with particle statistics imprint. Proceedings of the National Academy of Sciences, 119(21), p.e2119765119.

\bibitem {lee2022} Lee, D., Pramanik, T., Hong, S., Cho, Y.W., Lim, H.T., Chin, S. and Kim, Y.S., 2022. Entangling three identical particles via spatial overlap. Optics Express, 30(17), pp.30525-30535.

\bibitem {buhrman2001quantum} Harry Buhrman, Richard Cleve, John Watrous, and
Ronald De Wolf, “Quantum fingerprinting,” Physical Re-
view Letters 87, 167902 (2001).

\bibitem {barrett2007information} Jonathan Barrett, “Information processing in generalized
probabilistic theories,” Physical Review A 75, 032304
(2007).

\bibitem {plavala2021general} Plávala, M., 2023. General probabilistic theories: An introduction. Physics Reports, 1033, pp.1-64.

\bibitem {powell2008quantum} James R Powell, “The quantum limit to moore’s law,”
Proceedings of the IEEE 96, 1247–1248 (2008).

\bibitem {broadbent2013quantum} Anne Broadbent, Gus Gutoski, and Douglas Stebila,
“Quantum one-time programs,” in Annual Cryptology
Conference (Springer, 2013) pp. 344–360.

\bibitem {roehsner2018quantum} Marie-Christine Roehsner, Joshua A Kettlewell, Tiago B
Batalhao, Joseph F Fitzsimons, and Philip Walther,
“Quantum advantage for probabilistic one-time pro-
grams,” Nature communications 9, 1–8 (2018).


\end{thebibliography}

\begin{thebibliography}{10}

\bibitem {helstrom1969quantum} Carl W Helstrom, “Quantum detection and estimation theory,” Journal of Statistical Physics 1, 231–252 (1969).

\bibitem {peres1996separability} Peres, A., 1996. Separability criterion for density matrices. Physical Review Letters, 77(8), p.1413.

\end{thebibliography}
\end{document}